\documentclass[5p,times]{elsarticle}

\usepackage{array}
\usepackage{amsmath,amssymb,amsfonts}
\usepackage{algorithmic}
\usepackage{graphicx}
\usepackage{textcomp}
\usepackage{booktabs}
\usepackage{algorithm}
\usepackage{multirow}
\usepackage{threeparttable}
\usepackage{hyperref}
\usepackage{subfigure}
\usepackage{xcolor}  
\usepackage{tikz}  
\usetikzlibrary{arrows,shapes,chains} 

\usepackage{adjustbox}
\usepackage[font=small,labelfont=bf,labelsep=none]{caption} 
\biboptions{numbers,sort&compress}

\journal{Electric Power Systems Research}
\begin{document}

\begin{frontmatter}

\title{A Solution Strategy to the Unit Commitment Problem Incorporating Manifold Uncertainties\tnoteref{mytitlenote}}
\tnotetext[mytitlenote]{This work was supported by the National Natural Science Foundation of China under Grant 51777103.}


\author[mymainaddress]{Fang Zhai}

\author[mymainaddress]{Libao Shi\corref{mycorrespondingauthor}}
\cortext[mycorrespondingauthor]{Corresponding author}
\ead{shilb@sz.tsinghua.edu.cn}


\address[mymainaddress]{National Key Laboratory of Power System in Shenzhen, Shenzhen International Graduate School, Tsinghua University, Shenzhen 518055, China}

\begin{abstract}
The widespread uncertainties have made the interaction between wind power and power grid more complicated and difficult to model and handle. This paper proposes an approach for the solution of unit commitment (UC) problem incorporating multiple uncertainties that exist in both wind power and power grid inherently, consisting of probability, possibility, and interval measures. To handle the manifold uncertainties in a comprehensive and efficient manner, the evidence theory (ET) is applied to fuse these uncertain variables into Dempster-Shafer structure. Moreover, the power loss is introduced into power balance constraints, and the extended affine arithmetic (EAA) is employed to evaluate the uncertainty of power loss caused by the propagation of the aforementioned uncertainties. Regarding the mix-discrete nonlinear characteristics of the established optimization model, an enhanced grey wolf optimizer (GWO) algorithm is developed to solve the proposed model. Specifically, the corresponding commitment schedule is determined by a kind of binary grey wolf optimizer (BGWO), and the economic dispatch (ED) is settled by GWO. Finally, the IEEE 30-bus test system and a real-sized 183-bus China power system are studied to demonstrate the validity and scalability of the proposed model and method.
\end{abstract}


\end{frontmatter}


\section*{Nomenclature}
\label{sec:Nomenclature}
\subsection*{Acronyms}
\noindent\begin{tabular}{ll}
UC 		& Unit commitment. \\
SUC 	& Stochastic unit commitment. \\
RUC		& Robust unit commitment.\\ 
SO      & System operator. \\
ET      & Evidence theory. \\
DS      & Dempster-Shafer. \\
AA 	    & Affine arithmetic. \\
EAA     & Extended affine arithmetic. \\
P-box   & Probability box. \\
OPF     & Optimal power flow. \\
QF      & Quadratic form. \\
BGWO    & Binary grey wolf optimizer. \\
GWO     & Grey wolf optimizer. \\
SI      & Swarm intelligence.\\
\end{tabular}

\subsection*{Constants}
\noindent\begin{tabular}{ll} 
$T$			& Numbers of scheduling periods. \\
$NG$		& Set of thermal units.\\
$NB$		& Set of all buses. \\
$NW$		& Set of wind farms. \\
$\Lambda_{G}^{b}$ & Set of thermal units at bus $b$. \\
$\Lambda_{W}^{b}$ & Set of wind farms at bus $b$. \\
$C$ & Set of inequality constraints with DS structure.\\

\end{tabular}

\subsection*{Constants}
\noindent\begin{tabular}{ll}
$SU_{G(i)}^{t}$	& Startup cost of thermal unit $i$ at hour $t$. \\
$SD_{G(i)}^{t}$	& Shutdown cost of thermal unit $i$ at hour $t$. \\ 
$L_{G(i)}$  & Minimum generation capacity of unit $i$. \\
$U_{G(i)}$  & Maximum generation capacity of unit $i$. \\
$MUT_{G(i)}$ & Minimum up time of unit $i$. \\
$MDT_{G(i)}$ & Minimum down time of unit $i$. \\
$a_i$, $b_i$, $c_i$ &  Fuel cost coefficients of unit $i$. \\
$UR_{G(i)}$  &  Maximum ramp up rate of unit $i$. \\
$DR_{G(i)}$  &  Maximum ramp down rate of unit $i$. \\
$r_m$		& A pre-specified amount of the demand. \\
$K_{lk}^{b}$ & Line flow distribution factor for transmission line \\
 & to the net injection which links $l$ and $k$ \\
 & owing to the net injection at bus $b$.\\
$C_{lk}$	& Transmission capacity for transmission \\
& line linking $l$ and $k$. \\
$\overline{G}_{lk}$ & Conductance between bus $l$ and bus $k$.\\
$\overline{B}_{lk}$ & Susceptance between bus $l$ and bus $k$.\\
$\boldsymbol{\delta}$ & Matrix of phase angle difference.\\
$\boldsymbol{G}^{*}$ & Matrix composed of $\overline{G}_{lk}$.\\
$\boldsymbol{B}^{*}$ & Matrix composed of $\overline{B}_{lk}$.\\
$\boldsymbol{X}$ & Invertible matrix of $\boldsymbol{B}^{*}$.\\
\end{tabular}

\subsection*{Functions}
\noindent\begin{tabular}{ll}
$S_{c,G(i)}^{t}(\cdot)$		&  Transition cost function of thermal unit $i$ at hour $t$. \\
$F_{c,G(i)}^{t}(\cdot)$		&  Fuel cost function of thermal unit $i$ at hour $t$. \\
$G_c^t(\cdot)$			& Cost function from uncertain factors at hour $t$. \\
$P_{loss}^t(\cdot)$ &  Function of power loss at hour $t$. \\
$P_{d,b}^t(\cdot)$ &  Function of load demand at bus $b$ at hour $t$. \\
\end{tabular}

\subsection*{Variables}
\noindent\begin{tabular}{ll}
$t$			& Index of time intervals. \\
$i$			& Index of generation units. \\
$b,l,k$		& Index of buses. \\
$c$ & Index of constraints.\\
$u_{G(i)}^t$ & On/off(1/0) status of thermal unit $i$ at hour $t$.\\
$P_{G(i)}^t$ & Generation output of thermal unit $i$ at hour $t$. \\
\end{tabular}

\noindent\begin{tabular}{ll}
$\dot{P}_{W(i)}^{t}$ & Wind power output of unit $i$ \\
& modeled as probability distributions.  \\
$\tilde{P}_{d,b}^{t}$ & Uncertain load at bus $b$ at hour $t$ \\
& modeled as possible distributions.  \\
$\bar{\bar{P}}^{t}_{d,b}$ & Uncertain load at bus $b$ at hour $t$ \\
& modeled as interval measures.  \\
$P_{lk}^{t}$  & Power in transmission line linking $l$ and $k$ at hour $t$. \\
$SR_{G(i)}^t$ & Spinning reserve of thermal unit $i$ at hour $t$. \\
$\boldsymbol{\kappa}$ & Column vector of DS structures of all uncertain inputs.\\
$\boldsymbol{\kappa}_{0}$ & Column vector consisting of the range midpoints \\
& of the uncertain variables.\\
$\boldsymbol{\varepsilon}$ & Noise symbol. \\
$\hat{P}_{loss}$ & Quadratic form of power loss.\\
$P_{loss, 0}$ & Central value of power loss.\\
$\boldsymbol{P}$ & Matrix of power net injection.\\
$\hat{P}_{G(i)}^t$ & Quadratic form of $P_{G(i)}^t$.\\
$\hat{P}_{W(i)}^t$ & Quadratic form of $\dot{P}_{W(i)}^{t}$.\\
$\hat{P}_{d,b}^t$ & Quadratic form of $\tilde{P}_{d,b}^{t}$ and $\bar{\bar{P}}^{t}_{d,b}$.\\
$\hat{P}_{loss}^t$ & Quadratic form of $P_{loss}^t$.\\
$\hat{P}_{lk}^t$ & Quadratic form of $P_{lk}^t$.\\
$\Delta \hat{P}^t$ & Power imbalance at hour $t$.\\
$CV_c$  & Constraint violation in the $c$th constraint.\\
$TCV$   & Total constraint violation.\\
$P_{wt}$ & The output power of wind turbine.\\
\end{tabular}

\section{Introduction} 
\label{sec:introduction}
Unit commitment (UC), as one of the key applications of power generation scheduling, has been widely studied and utilized by system operators (SOs). Today in the background of fossil fuel depletion and serious environmental issues, how to find a low-cost and high-reliable solution of scheduling and dispatching generation units has become increasingly significant. However, there are many challenges and difficulties in the solution of UC problem, and one of the most common issues is how to effectively fuse the manifold uncertainties that widely exist in all aspects of power systems~\cite{zheng2015stochastic}. In particular, with the integration of large-scale renewable energy sources such as wind power and solar energy~\cite{tuohy2009unit,quan2015computational}, the interactions between uncertain renewable energy and power grid make power system operation more complicated and difficult to solve.    

For the sake of handling the uncertainties existed in the UC problem, researchers have proposed several modeling techniques. A survey of the literature indicated that the most widely used technique for the solution of UC was the probability related theory, which developed stochastic optimization in UC (also known as stochastic unit commitment, SUC)~\cite{tuohy2009unit,wang2008security,wu2007stochastic,pozo2013chance,wang2013stochastic}. Hence, numerical methods like Monte Carlo simulation (MCS) were adopted for scenario generation to approximate the uncertain factor distribution. In \cite{tuohy2009unit,wang2008security}, wind power scenarios were generated to describe the volatility and intermittency of wind generation. The uncertainties of load variation~\cite{pozo2013chance} and demand response~\cite{wang2013stochastic} were also considered in the same way. Although the scenario-based approach can provide relatively accurate results, it is difficult to evaluate the influence of the worst case resulting from the uncertainty factor, and a large number of scenarios may place an additional burden on computation. What's worse, stochastic optimization supported by precise probability distribution models required enormous amounts of historical data, which were hard to obtain in the realistic large-scale power system. Thus, robust optimization technique~\cite{ben2009robust} was proposed as an alternative modeling framework for uncertainty management and has gained substantial attention from SOs and researchers in recent years. Compared to stochastic optimization, robust optimization only needs moderate information, for example, the expected value and the variance of the uncertain variables, for constructing uncertainty set (usually representing as intervals). Consequently, robust unit commitment (RUC) can provide acceptable and useful generation scheduling and dispatch results in practice~\cite{bertsimas2013adaptive,street2011contingency,zhao2013multi,wang2017robust}. In terms of RUC research, apart from the uncertainty of wind power generation~\cite{wang2017robust,zhao2013multi}, the uncertainties resulting from nodal net injection~\cite{bertsimas2013adaptive}, unexpected unit and line outage \cite{street2011contingency, pozo2013chance,wang2016two}, unforeseen load fluctuation ~\cite{zhao2013multi, pozo2013chance, wang2013stochastic}  and open electricity market~\cite{dimitroulas2011new,sioshansi2009evaluating} were also taken into account. 

Unfortunately, there exist two obvious disadvantages in RUC researches. First, just using intervals to describe uncertain factors means the information deficiency. In other words, the level of information utilization is relatively low. Second, the optimization results based on RUC are usually too conservative, which leads to a relatively high operation cost for the realistic power system. For the former, the fuzzy set theory, as the representation of possible distributions, was adopted in UC problem to better interpret the limited information~\cite{el2004fuzzy,saber2006fuzzy,wang2016two}. For the latter, combining with multiple uncertainty modeling techniques has become attractive for UC researchers~\cite{wang2016two,zhao2013unified} to improve this over-conservative approach. In detail, a unified stochastic and robust unit UC model that takes advantage of both SUC and RUC was proposed in~\cite{zhao2013unified}. Wang~\cite{wang2016two} et al. presented a two-stage multi-objective UC model in which the probability distributions and fuzzy set theory were applied to modeling unit outage and load fluctuation, respectively. Moreover, some progress has already been made to treat multiple uncertainties in power system analysis. A kind of affine arithmetic (AA) method was developed for uncertain renewable generations in optimal power flow (OPF) problem~\cite{pirnia2014novel,vaccaro2017affine}, and weather-based OPF was correspondingly proposed based on AA~\cite{coletta2019affine}. As for incorporating probability, possible, and interval measures, Luo~\cite{luo2018uncertain, luo2018solution} et al. applied evidence theory (ET) to fuse hybrid uncertain factors on the solutions of uncertain power flow and optimal power flow. A similar approach has also been applied to optimal allocation of distributed generations problem~\cite{zhao2019multi}.

Besides, as a factor of 6\%-10\% of power generation according to the World Bank statistics~\cite{powerloss}, power loss should not be easily neglected in modern power system operation. The traditional method to calculate power loss was adding AC constraints in the UC model, and therefore, the Benders decomposition was developed to handle the nonlinear power flow equations~\cite{fu2005security,nasri2016network}. Despite the fact that the results of power loss based on the model with AC constraints were quite accurate, the corresponding high computational cost cannot satisfy the specific practical requirements. To mitigate the difficulty, a dynamic piecewise linear model for DC transmission losses for economic dispatch (ED) was presented~\cite{dos2011dynamic}. After that, Zhong~\cite{zhong2013dynamic} et al. developed a quadratically constrained quadratic program (QCQP) method to improve the accuracy based on DC constraints.  

It is worth noting that power loss was also affected by uncertain factors. To face the challenge of uncertainty propagation, the interval analysis (IA) method was proposed and can obtain the worst-case lower and upper bounds of variables~\cite{moore1966interval}. However, the interval-based method inherently assumed the independence of variables and ignores the possible correlation between the uncertain factors, often providing over-estimated results in practice. After that, AA theory was developed and has been applied to OPF problems in power systems~\cite{vaccaro2017affine}, which can take the correlation among variables into account and yield much tighter lower and upper bounds compared to IA~\cite{ding2015affine}. Recently, a solution of OPF incorporating wind power generation and grid uncertainties was proposed based on ET and extend affine arithmetic (EAA) framework as a meaningful effort~\cite{luo2018solution}. To the best of our knowledge, the researches about combining manifold uncertainties in solution of UC problem have not been explored, and how to further explore and exploit the interaction of manifold uncertain factors and the introduction of more complicated uncertain components (such as power loss) on the UC problem is worthy of a more in-depth research.  

In this paper, we propose an approach for solving day-ahead UC problem incorporating manifold uncertainties that are mainly due to volatile wind power generation and unforeseen load fluctuation. Multiple types of methods to model these uncertainties, consisting of probability distributions, possibility distributions, and interval measures, are employed simultaneously in the UC model. Along the way, SOs can fully take advantage of the available data and improve the level of information utilization. Aiming to deal with the manifold uncertainties in a realistic and effective manner, the evidence theory (ET) is applied to fuse these uncertain variables into Dempster-Shafer (DS) structure. This approach can acquire the best possible probability bounds for UC problems with different types of uncertainties, and the obtained results can alleviate the over-conservatism in RUC. The DC power flow and B-coefficient method are applied for computing power loss to achieve a better trade-off between efficacy and accuracy. Furthermore, EAA is introduced to evaluate the uncertainty of power loss caused by the propagation of the aforementioned uncertainties. Regarding the mix-discrete nonlinear optimization model, an enhanced grey wolf optimizer (GWO) algorithm is developed, in which the commitment schedule is determined by a kind of binary grey wolf optimizer (BGWO) and the economic dispatch (ED) is settled by GWO. Finally, the IEEE 30-bus test system and a real-sized 183-bus power grid in China are studied to demonstrate the validity and scalability of this research. 

The remainder of this paper is organized as follows. In Section 2, the mathematical formulation incorporating manifold uncertainties is addressed. Section 3 introduces an enhanced GWO algorithm to solve the proposed model. The numerical experiments are performed in Section 4. At last, Section 5 summarizes the conclusions. 


\section{Formulation of UC Problem with Manifold Uncertainties} 
\label{sec:Formulation of UC Problem with Manifold Uncertainties}
\begin{figure}[!ht]
\centering
\includegraphics[width=0.5\textwidth]{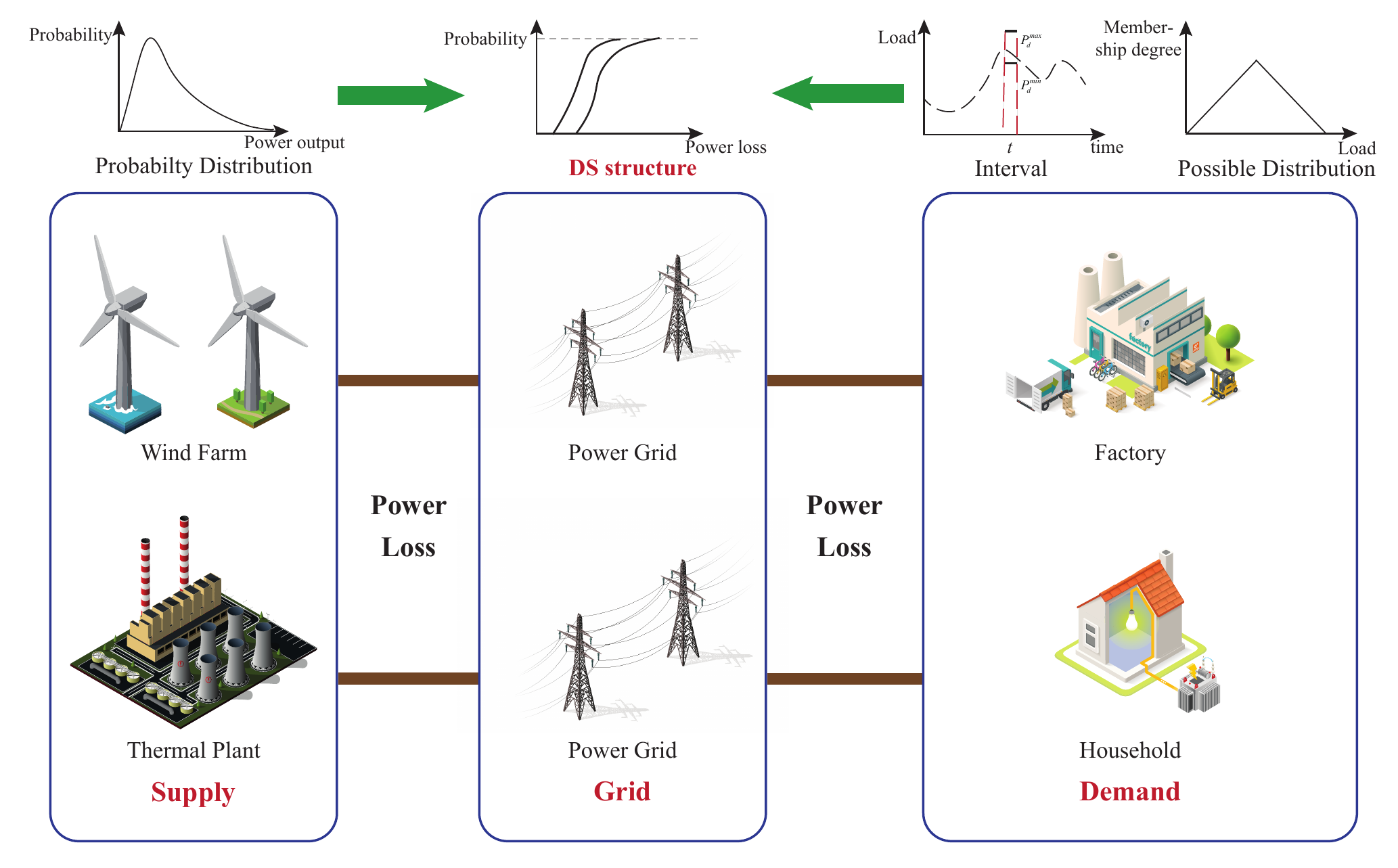}
\caption{Process of modeling construction.}
\label{fig-1}
\end{figure}

In this section, we first formulate the UC problem containing the uncertainties of wind power generation and load fluctuation. The wind power is modeled by probability distribution, and the load demands in different buses are described by intervals and possible distributions. Given that the wind power uncertainty is mainly homed by reserves of power grid in our work, the corresponding large-scale energy storage devices for wind farms are not taken into consideration during analysis. Apart from the traditional constraints, the uncertainty of power loss is also introduced into the power balance constraints. A kind of hybrid ET and EAA approach is correspondingly leveraged to treat the aforementioned manifold uncertainties reasonably. Finally, the mathematical model of UC problem incorporating manifold uncertainties is established. \autoref{fig-1} vividly illustrates the process of modeling construction.

\subsection{Problem formulation} 
\label{sub:problem_formulation}
In this paper, the objective function of UC problem is composed of the generation cost of thermal plants, including the fuel costs and the transition costs (i.e. start-up/shut-down costs), and the costs from the uncertainties. The operating cost of wind power generation is considered as zero over the short-term scheduling horizon. The corresponding expression of the objective function is shown as:
\begin{gather}
	\min \sum_{t=1}^{T} \sum_{i \in NG}[S_{c,G(i)}^{t}(u_{G(i)}^{t})+F_{c,G(i)}^{t}(u_{G(i)}^{t}, P_{G(i)}^{t})]+G_c^t(\cdot)  \\
	G_c^t(\cdot) = G_c^t(\dot{P}_{W(i)}^{t}, \tilde{P}_{d,b}^t, \bar{\bar{P}}^{t}_{d,b})
\end{gather}
where $G_c^t(\cdot)$ denotes the cost function from the uncertainties at hour $t$ (e.g. wind power generation and load fluctuation), $\dot{P}_{W(i)}^{t}$ is wind power generation of unit $i$ modeled by probability distribution, and $\tilde{P}_{d,b}^t$ and $\bar{\bar{P}}^{t}_{d,b}$ are the uncertain loads at bus $b$ at hour $t$ modeled by possible distributions and intervals, respectively. The detailed expression of $G_c^t(\cdot)$ is discussed in Section 2.2. Besides, The transition cost $S_{c,G(i)}^{t}(u_{G(i)}^{t})$ and fuel cost $F_{c,G(i)}^{t}(u_{G(i)}^{t}, P_{G(i)}^{t})$ are generally represented as:
\begin{align}
	& S_{c,G(i)}^{t}(u_{G(i)}^{t}) = SU_{G(i)}^{t}m_{G(i)}^{t}+SD_{G(i)}^{t}n_{G(i)}^{t} \\
	& m_{G(i)}^{t}=(1-u_{G(i)}^{t-1})u_{G(i)}^{t}, n_{G(i)}^{t}=(1-u_{G(i)}^{t})u_{G(i)}^{t-1} \\
	& F_{c,G(i)}^{t}(u_{G(i)}^{t}, P_{G(i)}^{t}) = u_{G(i)}^{t}[a_i(P_{G(i)}^{t})^2+b_iP_{G(i)}^{t}+c_i]
\end{align}	
where $a_i$, $b_i$, $c_i$ are fuel cost coefficients of unit $i$.

In fact, the objective function must be solved with the following constraints. These constraints include:
\begin{enumerate}[1)]
	\item Generation constraints:
	\begin{equation}
	u_{G(i)}^{t}L_{G(i)} \leq P_{G(i)}^{t} \leq u_{G(i)}^{t}U_{G(i)} \label{constraint-part1.1}
	\end{equation}
	where $L_{G(i)}$ and $U_{G(i)}$ are the minimum and maximum generation capacities of thermal unit $i$, respectively.
	\item Unit minimum uptime and downtime limitations:
	\begin{equation}
		T_{off,G(i)}^{t} \geq MDT_{G(i)}, T_{on, G(i)}^{t} \geq MUT_{G(i)}	
	\end{equation}
	where $T_{off,G(i)}^{t}$ and $T_{on, G(i)}^{t}$ represent the intervals that the unit $i$ has been shut down or started up at hour $t$.
	\item Ramp limitations:
	\begin{align}
	 	u_{G(i)}^{t}(P_{G(i)}^{t} - P_{G(i)}^{t-1}) \leq UR_{G(i)} \\
	 	u_{G(i)}^{t}(P_{G(i)}^{t} - P_{G(i)}^{t-1}) \geq -DR_{G(i)} \label{constraint-part1.2}
	\end{align} 
	\item Power balance constraints:
	\begin{gather}
		\sum_{i\in NG}P_{G(i)}^t+\sum_{i\in NW}\dot{P}_{W(i)}^t-P_{loss}^t(\cdot)=\sum_{b\in NB}P_{d,b}^t(\cdot)\\
		P_{loss}^t(\cdot) = P_{loss}^{t}(P_{G(i)}^{t}, \dot{P}_{W(i)}^{t}, \tilde{P}_{d,b}^t, \bar{\bar{P}}^{t}_{d,b}) \\
		P_{d,b}^t(\cdot) = P_{d,b}^t(\tilde{P}_{d,b}^t, \bar{\bar{P}}^{t}_{d,b})
	\end{gather}
	where $P_{loss}^t(\cdot)$ means the function of power loss at hour $t$ in the whole power system, and $P_{d}^t(\cdot)$ means the function of load at bus $b$ at hour $t$.
	Actually, how to formulate $P_{loss}^t(\cdot)$ and $P_{d}^t(\cdot)$ is one of key issues in our research, and further discussion can be found in the following sections.
	\item Spinning reserve limitations:
	\begin{gather}
		\sum_{b\in NB}(1+r_m)P_{d,b}^{t}(\tilde{P}_{d,b}^t, \bar{\bar{P}}^{t}_{d,b}) \leq \sum_{i\in NG} SR_{G(i)}^t \\
		SR_{G(i)}^t = \min \big\{u_{G(i)}^{t}(U_{G(i)}^{t}-P_{G(i)}^{t}), u_{G(i)}^{t}RR_{G(i)}\big\} \label{constraint-part3.2}
	\end{gather}
	where  $r_m$ is usually defined as a pre-specified amount of the load demand.
	\item Network security constraints:
	\begin{gather}
	P_{lk}^{t} = \sum_{b\in NB} K_{lk}^b \big[\sum_{i \in \Lambda_{G}^{b}} P_{G(i)}^{t} + \sum_{i \in \Lambda_{W}^{b}} \dot{P}_{W(i)}^{t} - P_{d,b}^{t}(\cdot) \big]\\
	-C_{lk} \leq P_{lk}^{t} \leq C_{lk}
	\end{gather}
	where $P_{lk}^{t}$ means active power in transmission line linking bus $l$ and bus $k$ at hour $t$. The network security constraints indicate the limitations of transmission capacity constraints~\cite{wang1995short}. 
\end{enumerate}


\subsection{Methods to handle manifold uncertainties} 
\label{sub:methods_to_handle_manifold_uncertainties}
\subsubsection{The hybrid ET and EAA approach}
ET is a useful and powerful approach to combine various types of uncertain information from different sources. In ET, DS structure is applied to represent uncertainty, which is capable of describing both aleatory and epistemic uncertainty attribute of variables. Specifically, a mass function $m$, also called basic probability assignment (BPA), satisfies the requirements as given in~(\ref{def-1}) and defines the belief and plausibility measures as given in~(\ref{def-2})~\cite{zio2013literature}.
\begin{gather}
	m: 2^{\Omega} \to [0,1], m(\varnothing) = 0, \sum_{A \in 2^{\Omega}}m(A)=1 \label{def-1}\\
	Bel(A) = \sum_{B \subseteq A} m(B), Pl(A) = \sum_{B \cap A \neq 0} m(B) \label{def-2}
\end{gather}
where ${\Omega}$ means the universe, and $2^{\Omega}$ denotes the power set of all subsets of $\Omega$. The above measures limit the probability P of any set $A \in 2^{\Omega}$ as given in~(\ref{def-3}) and therefore, DS structure can transform into a probability box (P-box) equivalently, and vice versa~\cite{ferson2015constructing}. 
\begin{equation}
	Bel(A) \leq P(A) \leq Pl(A) \label{def-3}
\end{equation}

Based on the ET framework, the probability distributions, possible distributions, and intervals can be converted into DS structures with a finite number $n_{X}$ of closed intervals~\cite{ferson2015constructing,baudrit2006joint}. As to the binary arithmetic operation $\Box$ with two independent variables $X$ ($X=\{(x_i,m_X(x_i)|i=1,...,n_X)\}$) and $Y$ ($Y=\{(y_i,m_Y(y_i)|i=1,...,n_Y)\}$), the DS result $Z=X\Box Y$ ($Z=\{(z_{ij},m_Z(z_{ij})|i=1,...,n_X,j=1,...,n_Y)\}$) is calculated by $z_{ij}=x_i\Box y_j, m_Z(z_{ij})=m_X(x_i)m_Y(y_i)$. Once considering the dependency between variables, if the P-boxes of X, Y, and Z are denoted by $[\overline{F}_X,\underline{F}_X]$, $[\overline{F}_Y,\underline{F}_Y]$, and $[\overline{F}_Z,\underline{F}_Z]$, the convolutions under perfect, opposite, and unknown dependence are expressed as~(\ref{def-add-1}), (\ref{def-add-2}) and~(\ref{def-add-3})\ \cite{luo2018solution,oberkampf2004dependence}.
\begin{equation}
\left\{
\begin{aligned}
	\overline{F}^{-1}_Z(p)=\inf [\overline{F}^{-1}_X(p)\Box \overline{F}^{-1}_Y(p)], F_X \in [\overline{F}_X,\underline{F}_X], F_Y \in [\overline{F}_Y,\underline{F}_Y]\\
	\overline{F}^{-1}_Z(p)=\sup [\overline{F}^{-1}_X(p)\Box \overline{F}^{-1}_Y(p)], F_X \in [\overline{F}_X,\underline{F}_X], F_Y \in [\overline{F}_Y,\underline{F}_Y]\\
\end{aligned}	
\right.
\label{def-add-1}
\end{equation}
\begin{equation}
\left\{
\begin{aligned}
	\overline{F}^{-1}_Z(p)=\inf [\overline{F}^{-1}_X(1-p)\Box \overline{F}^{-1}_Y(p)], 
	\overline{F}_X \in [\overline{F}_X,\underline{F}_X], \overline{F}_Y \in [\overline{F}_Y,\underline{F}_Y]\\
	\overline{F}^{-1}_Z(p)=\sup [\overline{F}^{-1}_X(1-p)\Box \overline{F}^{-1}_Y(p)],
	\overline{F}_X \in [\overline{F}_X,\underline{F}_X], \overline{F}_Y \in [\overline{F}_Y,\underline{F}_Y]\\
\end{aligned}	
\right.
\label{def-add-2}
\end{equation}
\begin{equation}
\left\{
\begin{aligned}
	& \overline{F}^{-1}_Z(z)=\inf \min[F_X(x)+F_Y(y),1], \\
	& z=x\Box y, F_X \in [\overline{F}_X,\underline{F}_X], F_Y \in [\overline{F}_Y,\underline{F}_Y]\\
	& \underline{F}^{-1}_Z(z)=\sup \max[F_X(x)+F_Y(y)-1,0],\\
	& z=x\Box y, F_X \in [\overline{F}_X,\underline{F}_X], F_Y \in [\overline{F}_Y,\underline{F}_Y]\\
\end{aligned}	
\right.
\label{def-add-3}
\end{equation}
Moreover, we apply EAA to calculate the effect of uncertainty propagation with DS structure. Comparing to AA, EAA is able to record the first and second order correlations between variables and noise symbols. A variable formulated by a quadratic form (QF) with DS noise symbols is represented as follows:
\begin{equation}
	\hat{x} = x_0 + \boldsymbol{X}_1\boldsymbol{\varepsilon} + \boldsymbol{\varepsilon}^T\boldsymbol{X}_2\boldsymbol{\varepsilon} \label{def-4}
\end{equation}
where $x_0$ is the central value, and $\boldsymbol{\varepsilon}$ is the column vector of DS noise symbols. $\boldsymbol{X}_1$ and $\boldsymbol{X}_2$ are the matrices of partial deviations, which indicate the first and second order correlations with $\boldsymbol{\varepsilon}$. Arithmetic operations on QFs are summarized as~(\ref{def-5}). Thus, a QF with DS noise symbols can be converted into a DS structure and denoted by $DS(\hat{x})$.
\begin{equation}
\left\{
\begin{aligned}
	\alpha \pm \hat{x} = & (a \pm x_0) \pm \boldsymbol{X}_1\boldsymbol{\varepsilon}\pm  \boldsymbol{\varepsilon}^T\boldsymbol{X}_2\boldsymbol{\varepsilon},\alpha \in R \\
	\alpha \cdot \hat{x} = & \alpha x_0+(\alpha\boldsymbol{X}_1)\boldsymbol{\varepsilon}+\boldsymbol{\varepsilon}^T(\alpha \boldsymbol{X}_2)\boldsymbol{\varepsilon},\alpha \in R \\
	\hat{x}\pm \hat{y} = & (x_0 \pm y_0)+(\boldsymbol{X}_1\pm\boldsymbol{Y}_1)\boldsymbol{\varepsilon} + \boldsymbol{\varepsilon}^T(\boldsymbol{X}_2\pm\boldsymbol{Y}_2)\boldsymbol{\varepsilon} \\
	\hat{x}\cdot\hat{y} = & x_0y_0+(y_0\boldsymbol{X}_1+x_0\boldsymbol{Y}_1)\boldsymbol{\varepsilon} \\& +\boldsymbol{\varepsilon}^T(y_0\boldsymbol{X}_2+x_0\boldsymbol{Y}_2+\boldsymbol{X}_1^T\boldsymbol{Y}_1)\boldsymbol{\varepsilon}
\end{aligned}	
\right.
\label{def-5}
\end{equation}

\subsubsection{The uncertainty of power loss}
The B-coefficient method has proven to be an efficient and powerful tool to rapidly calculate power loss with high enough accuracy~\cite{dos2011dynamic,wood2013power,aoki1982economic}. If we assume that $\Delta \theta_{lk}$ is the phase angle difference between bus $l$ and bus $k$, and under the approximation $\cos \Delta \theta_{lk} \approx 1-\frac{1}{2}\Delta \theta_{lk}^2$, the power loss can be represented as:
\begin{gather}
	P_{loss} = \boldsymbol{\delta}^T \boldsymbol{G}^{*} \boldsymbol{\delta}+ P_{LV}, P_{LV} = \sum_{l\in NB} \sum_{k\in NB} (U_l - U_k)^2\overline{G}_{lk} \label{equ-ploss-1}
\end{gather}
where $\boldsymbol{\delta}$ is the phase angle difference matrix. $U_l$ and $U_k$ represent the voltage magnitudes of bus $l$ and bus $k$, respectively, and $\overline{G}_{lk}$ represents the conductance between bus $l$ and bus $k$. The elements in the matrix of $\boldsymbol{G}^{*}$ are expressed as:
\begin{equation}
	G^{*}_{ll} = \sum_{k\in NB} U_lU_k\overline{G}_{lk}, G^{*}_{lk} = -U_lU_k\overline{G}_{lk} (l \neq k) \\
\end{equation}
According to the idea of DC power flow, the relationship between the net power injection $\boldsymbol{P}$ and the phase angle difference $\boldsymbol{\delta}$ is formulated as:
\begin{equation}
	\boldsymbol{P} = \boldsymbol{B}^{*}\boldsymbol{\delta}, \boldsymbol{XP} = \boldsymbol{\delta}, \boldsymbol{X} = [\boldsymbol{B}^{*}]^{-1}
\end{equation}
where the elements in the matrix $\boldsymbol{B}^{*}$ are given in (\ref{equ-ploss-2}), and $\overline{B}_{lk}$ means the susceptance between bus $l$ and bus $k$.
\begin{equation}
	B_{ll}^{*} = -\sum_{k\in NB} U_lU_k\overline{B}_{lk}, B_{lk}^{*} = U_lU_k\overline{B}_{lk} (l\neq k) \label{equ-ploss-2}
\end{equation}
Substituting (\ref{equ-ploss-2}) into (\ref{equ-ploss-1}), the system power loss can be finally expressed as:
\begin{equation}
	P_{loss} = \boldsymbol{P}^T\boldsymbol{X}^T\boldsymbol{G}^{*}\boldsymbol{X}\boldsymbol{P} + P_{LV} \label{def-ploss}
\end{equation}

Next, we will apply the hybrid ET and EAA approach to formulate the uncertainty of system power loss. If the column vector $\boldsymbol{\kappa}$ denotes the DS structure of all uncertain inputs ($\dot{P}_{W(i)}^{t}$, $\tilde{P}_{d,b}^t$, $\bar{\bar{P}}^{t}_{d,b}$), the noise symbol $\boldsymbol{\varepsilon}$ is represented as (\ref{def-noise}) through the following normalization:
\begin{equation}
	\boldsymbol{\varepsilon} = \Delta\boldsymbol{\kappa}^{-1}(\boldsymbol{\kappa}-\boldsymbol{\kappa}_{0}) \label{def-noise}
\end{equation}
where $\boldsymbol{\kappa}_{0}$ is the column vector consisting of the range midpoints of the uncertain variables, and $\Delta\boldsymbol{\kappa}$ is the diagonal matrix of the range radiuses  of the variables~\cite{luo2018solution}. Based on the method of EAA, the QFs of power loss $\hat{P}_{loss}$ can be expressed as:
\begin{equation}
	\hat{P}_{loss} = P_{loss, 0} + \frac{\partial P_{loss}}{\partial \boldsymbol{\kappa}} \Big|_{\boldsymbol{\kappa}_0}\Delta \boldsymbol{\kappa}\boldsymbol{\varepsilon}+\boldsymbol{\varepsilon}^T \frac{1}{2}  \Delta \boldsymbol{\kappa} \frac{\partial^2 P_{loss}}{\partial \boldsymbol{\kappa}^2} \Big|_{\boldsymbol{\kappa}_0} \Delta \boldsymbol{\kappa}\boldsymbol{\varepsilon}
\end{equation}
where $P_{loss, 0}$ is the central value of power loss, which can be obtained by (\ref{def-ploss}) at the central value of uncertain inputs. Besides, we can calculate the partial derivatives according to (\ref{def-ploss}), as shown by:
\begin{gather}
 	\frac{\partial P_{loss}}{\partial \boldsymbol{\kappa}} = [\boldsymbol{X}^T\boldsymbol{G}^{*}\boldsymbol{X} + (\boldsymbol{X}^T\boldsymbol{G}^{*}\boldsymbol{X})^T]\boldsymbol{P}\label{def-ploss-partial-1}\\
 	\frac{\partial^2 P_{loss}}{\partial \boldsymbol{\kappa}^2} = \boldsymbol{X}^T\boldsymbol{G}^{*}\boldsymbol{X} + (\boldsymbol{X}^T\boldsymbol{G}^{*}\boldsymbol{X})^T
\label{def-ploss-partial-2}
\end{gather}
Hence, $\frac{\partial P_{loss}}{\partial \boldsymbol{\kappa}}$ and $\frac{\partial^2 P_{loss}}{\partial \boldsymbol{\kappa}^2}$ are the submatrices of (\ref{def-ploss-partial-1}) and (\ref{def-ploss-partial-2}) composed of the positions of corresponding uncertain variables.

\subsubsection{Treatment of constraints} 
With the framework of ET and EAA, the power balance constraints, spinning reserve limitations, and network security constraints can be modified as the equations with DS structure. However, regarding the existence of manifold uncertainties, the equations with DS structure cannot be strictly satisfied. Inspired by the methodology of chance constraint, power balance constraints are modified as follows:
\begin{equation}
	\Delta \hat{P}^t=\sum_{i\in NG} \hat{P}_{G(i)}^t+\sum_{i \in NW}\hat{P}_{W(i)}^t-\hat{P}_{loss}^t-\sum_{b \in NB} \hat{P}_{d,b}^t \label{constraint-part2.1}
\end{equation}
where $\hat{P}_{G(i)}^t$, $\hat{P}_{W(i)}^t$, $\hat{P}_{loss}^t$, and $\hat{P}_{d,b}^t$ are the QFs of corresponding variables, and $\Delta \hat{P}^t$ means power imbalance at time $t$. Let $\Delta$ denote the tolerance level of power imbalance, which is usually set as a small amount of the demand, and therefore, $\Delta \hat{P}^t$ satisfies the following constraint:
\begin{equation}
	- \Delta \le DS(\Delta \hat{P}^t) \le \Delta \label{constraint-part2.2}
\end{equation}
It is worth noting that there exists certain errors in the calculation of DC power flow. Accordingly, setting $\Delta$ as the tolerance level of power imbalance is practical for SOs to schedule generation units. After introducing the concept of $\Delta$, the cost function caused by uncertainties $G_c(\cdot)$ in the objective function is expressed as:
\begin{equation}
	G_c^t(\dot{P}_{W(i)}^{t}, \tilde{P}_{d,b}^t, \bar{\bar{P}}^{t}_{d,b}) = \xi_{p} (\Delta \hat{P}^t)^2
\end{equation}
where $\xi_{p}$ means the penalty cost owing to power imbalance.

Second, applying the same method, the limitations of the spinning reserves can be represented as:
\begin{equation}
	DS\left((1+r_m)\sum_{b \in NB} \hat{P}_{d,b}^t\right) \leq \sum_{i\in NG} SR_{G(i)}^t \label{constraint-part3.1}
\end{equation}
and the network security constraints are converted to the following forms:
\begin{gather}
	\hat{P}_{lk}^t = \sum_{b \in B} K_{lk}^b [\sum_{i \in \Lambda_{G}^{b}} \hat{P}_{G(i)}^{t} + \sum_{i \in \Lambda_{W}^{b}} \hat{P}_{W(i)}^{t} - \hat{P}_{d,b}^t] \label{constraint-part4.1}\\
	-C_l \leq DS\left(\hat{P}_{lk}^b\right) \leq C_l \label{constraint-part4.2}
\end{gather}
where $\hat{P}_{lk}^t$ means the QF of ${P}_{lk}^t$.

\subsection{UC model with DS structure} 
\label{sub:uc_model_with_ds_structure}
The ultimate UC model incorporating manifold uncertainties with DS structure can be described as follows:
\begin{equation*}
\begin{aligned}
	\min DS & \left(\sum_{t=1}^{T} \sum_{i\in NG}\left[S_{c,G(i)}^{t}(u_{G(i)}^{t})+F_{c,G(i)}^{t}(u_{G(i)}^{t}, P_{G(i)}^{t})\right]+\xi_{p} \Delta (\hat{P}^t)^2 \right) \\
	& \begin{aligned}
	& s.t. &&(\ref{constraint-part1.1})-(\ref{constraint-part1.2}), \forall i\in NG,\forall t \\
	& && (\ref{constraint-part3.2}), (\ref{constraint-part2.1}), (\ref{constraint-part2.2}), (\ref{constraint-part3.1}),  \forall t\\
	& && (\ref{constraint-part4.1}), (\ref{constraint-part4.2}), \forall l\in NB, \forall k \in NB, \forall t	
	\end{aligned}
\end{aligned}
\end{equation*}

\section{Solution algorithm} 
\label{sec:solution_algorithm}
Researchers have proposed various algorithms for the solution of UC problems, including traditional mathematical programming methods and computational intelligence methods. For the established model, it is difficult to express the process of converting manifold uncertainties into DS structures, and to compare fitness values with DS structures (as shown in Section~\ref{fitness_comparison}) by using traditional mathematical programming techniques. Besides, to handle these uncertainties and evaluate the uncertainty of power loss, many nonlinear equations and constraints are specifically introduced. Regarding that the proposed UC model with DS structure bears complicated, mix-integer and nonlinear characteristics, an enhanced GWO algorithm is developed for the solution of the proposed model.

\subsection{An enhanced GWO algorithm} 
\label{sub:an_enhanced_bgwo_and_GWO_algorithm}
GWO, as one of the recent swarm intelligence based algorithms, was initially proposed by Mirjalili~\cite{mirjalili2014grey} et al. in 2014. Inspired by the leadership hierarchy and hunting mechanisms of grey wolves, the wolf pack is divided into four groups in the GWO algorithm, namely, the alpha wolf ($\alpha$), the beta wolf ($\beta$), the delta wolf ($\delta$), and the omega wolf ($\omega$). The mathematical expressions of wolf behaviors to search the best solution are summarized as follows~\cite{mirjalili2014grey, faris2018grey}:
\begin{gather}
	\vec{D} = |\vec{C} \cdot \vec{X}_{p}(iter) - \vec{X}(iter)|,
	\vec{X}(iter+1) = \vec{X}_{p}(iter) - \vec{A} \cdot \vec{D} \\
	\vec{A} = 2\vec{a}\cdot \vec{r}_1 - \vec{a}, \vec{C} = 2\cdot \vec{r}_2, a = 2-iter*2/MaxIter\\
	\left\{
\begin{aligned}
\vec{D}_{\alpha} &= |\vec{C}_1 \cdot \vec{X}_{\alpha} - \vec{X}|, \ \vec{X}_1 = \vec{X}_{\alpha} - \vec{A}_1 \cdot(\vec{D}_{\alpha})\\
\vec{D}_{\beta} &= |\vec{C}_2 \cdot \vec{X}_{\beta} - \vec{X}|, \ \vec{X}_2 = \vec{X}_{\beta} - \vec{A}_2 \cdot(\vec{D}_{\beta}) \\
\vec{D}_{\delta} &= |\vec{C}_3 \cdot \vec{X}_{\delta} - \vec{X}|, \ \vec{X}_3 = \vec{X}_{\delta} - \vec{A}_3 \cdot(\vec{D}_{\delta}) 		
\end{aligned}	
\right.\label{GWO-equ-1}\\
\vec{X}(iter+1) = (\vec{X}_1+\vec{X}_2+\vec{X}_3)/3\label{GWO-equ-2}
\end{gather}
where $iter$ is the number of iteration, $\vec{A}$ and $\vec{C}$ are the vectors of coefficients, $\vec{X}_p$ is the vector of prey position, $\vec{X}$ is the vector of grey wolf position near prey, each element in $\vec{r}_1$ and $\vec{r}_2$ is the random number in [0,1], $MaxIter$ is the maximum iteration number, $\vec{X}_{\alpha}$, $\vec{X}_{\beta}$ and $\vec{X}_{\delta}$ denote the positions of the alpha wolf, the beta wolf and the delta wolf, respectively, and $\vec{X}(iter+1)$ is the vector of grey wolf position in the next iteration.

In comparison to other metaheuristic swarm intelligence approaches, GWO has two interesting characteristics~\cite{faris2018grey}:1) it is easy to leverage fewer parameters to adjust the optimization process during iteration, which relieves the burden of finding the best parameter; 2) it strikes a balance between exploration and exploitation during searching better solution shown by the parameter $a$, leading to a preferable convergence. In this paper, we adopt the method of decreasing parameter $a$ with an exponential function ($a=2*(1-iter^2/MaxIter^2)$), to extend the process of exploration during iteration~\cite{mittal2016modified}. 

The original GWO algorithm is developed to optimize continuous real-valued functions and hard to find the optimal solution of the problem including binary variables. Thus, some researchers have developed BGWO for binary optimization~\cite{emary2016binary,panwar2018binary}, which are described as follows:
\begin{align}
\vec{B}(iter+1) & = \begin{cases}
1 & if \ S\left(\vec{X}(iter+1)\right) > r \\
0 & otherwise
\end{cases} \\
S\left(\vec{X}(iter+1)\right) & = 1/(1+e^{-10\{(\dfrac{\vec{X}_1+\vec{X}_2+\vec{X}_3}{3}-0.5)\}})	
\end{align}
where $t$ is the number of iteration, $\vec{B}$ is the vector of binary variables, $r$ is the random number obeying uniform distribution $U[0,1]$, and the function $S(\cdot)$ is actually the sigmoid transformation function.

\subsection{Key Treatments to UC solution} 
\label{sub:bgwo_and_gwo_algorithms_application_to_uc_problem}
Briefly, the aforementioned BGWO algorithm is applied for the unit commitment schedule in UC problem, and the enhanced GWO algorithm is used for ED problem. In this section, we introduce some key treatments, containing fitness comparison, inequality constraints with DS structure, initial population optimization and constraints repair, for applying the GWO algorithm to UC problems with manifold uncertainties.

\subsubsection{Fitness comparison}
\label{fitness_comparison}
When a fitness value is set as DS structure, how to compare the two DS fitness values is worth making further discussion. In~\cite{compare2015genetic}, a ranking method in ET using probability bounds is proposed, but this method cannot distinguish fitness values with large overlaps in probability bounds. Instead, we adopt a quantile-based method in this paper to compare the fitness~\cite{luo2018solution}, which can avoid poor optimization results because of considering plenty of fitness as equal. Suppose that $\alpha$ and $\beta$ are the fitness values, and $Q_{\alpha}$ and $Q_{\beta}$ are their corresponding quantile functions. If $\alpha$ is better than $\beta$, the following statement should be satisfied:
\begin{equation}
	(\forall p \in [0,1], Q_{\alpha}(p) \le Q_{\beta}(p)) \wedge (\exists p_0 \in [0,1], Q_{\alpha}(p_0) \neq Q_{\beta}(p_0))
\end{equation}
which means that in P-box, the left bound of better fitness value is better than the others, and the right bound is at least equal, or the left bound of better fitness value is equal to the others, but the right bound is better. 

\subsubsection{Treatment for inequality constraints with DS structure}
Regarding the inequality constraints with DS structure existing in our proposed model, the concepts of lower and upper probabilities of a variable $m$ satisfying an inequality constraint in P-box are introduced to handle these constraints. Let $\underline{F}_{M}$ and $\overline{F}_{M}$ represent the cumulative probability functions (CDFs) of the left and right bounds of $M$, and the requirements of meeting inequality constraints are listed as follows~\cite{luo2018solution}:
\begin{equation}
\begin{cases}
\underline{F}_{M}(m) \le P(M \le m) \le \overline{F}_{M} \\
1-\overline{F}_{M}(m) \le P(M \ge m) \le 1-\underline{F}_{M}(m)
\end{cases}
\end{equation}

Suppose that $X$ is the column vector of decision variables, and $g_c(X)$ is the $c$th inequality constraint with DS structure. Then, the inequality constraint $c$ ($c\in C$) can be treated as a chance constraint, i.e. 1) if the lower probability is not less than the pre-set threshold $\sigma_c$, the constraint is satisfied; 2) otherwise, calculate the constraint violation ($CV_c$):
\begin{equation}
	CV_c=\max\left\{\sigma_c-\inf Pr[g_c(X)\le 0],0\right\}
\end{equation}

Therefore, there are two cases when a solution is more preferable than the other: 1) the total constraint violation ($TCV$, $TCV=\sum_{t=1}^{T}\sum_{c\in C} CV_c$) is less than the other; 2) if $TCV$ is the same, the fitness value is better.

\subsubsection{Initial population optimization and constraints repair}
The classical GWO algorithm just generates initial population randomly. Inspired by~\cite{wang2016two,wang2013supply}, we introduce priority list method to optimize the initial population and repair constraints during initialization and iterations. The details are summarized as follows:

\textbf{[Step 1]} Initialization: The initial values of $u_{G(i)}^{t}$ and $P_{G(i)}^{t}$ are generated randomly, where $u_{G(i)}^{t}$ is either 0 or 1, and $P_{G(i)}^{t}$ is within the range of $[L_{G(i)}, U_{G(i)}]$.

\textbf{[Step 2]} Priority list to adjust $u_{G(i)}^{t}$: Calculate the priority coefficient $\lambda_i$ of each unit as given in (\ref{equ-calculate-lambda}), and define $PL$ as the set of all units ascending sort by $\lambda_i$ (the first element in $PL$ is denoted as $G(a)$).
\begin{equation}
	\lambda_{i} =  F_c(U_{G(i)})/U_{G(i)}=a_i/U_{G(i)}+b_i+c_i*U_{G(i)} \label{equ-calculate-lambda}
\end{equation}
For each $t$, $\Delta P^{t} = \sum_{i\in NG}u_{G(i)}^{t}U_{G(i)}-\sum_{b\in NB}P_{d,b}^t$. If $\Delta P^{t} < 0$, set an off-line unit $u_{G(i)}^{t}$ with the lowest $\lambda_i$ as 1 until $\Delta P^{t} \ge 0$. Next, a heuristic-based constraint treatment for minimum uptime and downtime limitations is adopted~\cite{panwar2018binary,srikanth2018meta}, as shown in Algorithm 1.

However, the capacity based on this treatment may over-supply compared to demand. Thus, let $DPL$ be the set of all units descending sort by $\lambda_i$; denote the first element in $DPL$ is $G(b)$, and calculate $\Delta P^{t} = \sum_{i\in NG}u_{G(i)}^{t}U_{G(i)}-P_{d}^t$. If $\Delta P^{t} < U_{G(b)}$ or $\Delta P^{t} \ge U_{G(b)}$ with the violation of minimum on/off time constraints, delete $G(b)$ from $DPL$. Otherwise, set $u_{G(b)}^{t}=1$ and delete $G(b)$ from $DPL$. The loop is continued until $DPL=\varnothing$.
\begin{algorithm} 
\caption{Pseudo code for minimum up/down time limitations treatment} 
\label{table-pseudo-1} 
\begin{algorithmic} 
\STATE \textbf{Begin}
\STATE \quad \textbf{For} $i\in NG$
\STATE \quad \quad \textbf{If} unit $i$ is set to be on at hour $t$ (i.e., $u_{G(i)}^{t}=1$),\textbf{then}
\STATE \quad \quad \quad \quad \textbf{If} $T_{off,G(i)}^{t-1}<MDT_{G(i)}$, \textbf{then} $u_{G(i)}^{t}=0$
\STATE \quad \quad \quad \quad \textbf{Else if} $T_{off,G(i)}^{t-1} \ge MDT_{G(i)}$, \textbf{then} $u_{G(i)}^{t}=1$
\STATE \quad \quad \quad \quad \textbf{End if}
\STATE \quad \quad \quad \textbf{Else if} $u_{G(i)}^{t-1}=1$, \textbf{then} $u_{G(i)}^{t}=1$
\STATE \quad \quad \quad \textbf{End if}
\STATE \quad \quad \textbf{Else if} $u_{G(i)}^{t}=0$, \textbf{then}
\STATE \quad \quad \quad \textbf{If} $u_{G(i)}^{t-1}=1$, \textbf{then}
\STATE \quad \quad \quad \quad \textbf{If} $T_{on,G(i)}^{t-1}<MUT_{G(i)}$, \textbf{then} $u_{G(i)}^{t}=1$
\STATE \quad \quad \quad \quad \textbf{Else if} $T_{on, G(i)}^{t-1} \ge MUT_{G(i)}$, \textbf{then} $u_{G(i)}^{t}=0$
\STATE \quad \quad \quad \quad \textbf{End if}
\STATE \quad \quad \quad \textbf{Else if} $u_{G(i)}^{t-1}=0$, \textbf{then} $u_{G(i)}^{t}=0$
\STATE \quad \quad \quad \textbf{End if}
\STATE \quad \quad \textbf{End if}
\STATE \quad \textbf{End for}
\STATE \textbf{End}
\end{algorithmic}
\end{algorithm}

\textbf{[Step 3]} Adjustment for ED problem: Calculate $\Delta P^{t} = \sum_{b\in NB} P_{d,b}^t-\sum_{i\in NG}u_{G(i)}^{t}P_{G(i)}^t$. If $u_{G(a)}^{t}=0$, delete $G(a)$ from $PL$. If $u_{G(a)}^{t}=1$, calculate $\Delta P_{G(a)}^t = P_{G(a)}^{max}-P_{G(a)}^t$. Furthermore, if $\Delta P_{G(a)}^t \ge \Delta P^{t}$, set $P_{G(a)}^t=P_{G(a)}^t+\Delta P^{t}$ and delete $G(a)$ from $PL$. Otherwise, $P_{G(a)}^t=P_{G(a)}^t+rand(U_{G(a)}-P_{G(a)}^{t})$ and delete $G(a)$ from $PL$. Next, recalculate $\Delta P^{t}$. The loop will be stopped when $PL=\varnothing$.

After the dispatch adjustment, the ramp limitations are also required. Similarly, Algorithm 2 summarizes the treatment.
\begin{algorithm} 
\caption{Pseudo code for ramp limitations treatment} 
\label{table-pseudo-2} 
\begin{algorithmic} 
\STATE \textbf{Begin}
\STATE \quad \textbf{For} $i\in NG$
\STATE \quad \quad \textbf{For} $t=2$ to MaxHour 
\STATE \quad \quad \quad $\Delta P_{G(i)}^{t}=P_{G(i)}^{t}-P_{G(i)}^{t-1}$ 
\STATE \quad \quad \quad \textbf{If} $\Delta P_{G(i)}^{t}<-DR_{G(i)}$, \textbf{then} 
\STATE \quad \quad \quad \quad $P_{G(j)}^{t}=P_{G(j)}^{t-1}-DR_{G(i)}$, $\Delta D_{i} = -DR_{G(i)}-\Delta P_{G(i)}^t$.
\STATE \quad \quad \quad \quad Dispatch $\Delta D_{i}$ in the order of $DPL$
\STATE \quad \quad \quad \quad subject to ramp and generation constraints.
\STATE \quad \quad \quad \textbf{Else if} $\Delta P_{G(i)}^{t} > UR_{G(i)}$, \textbf{then} 
\STATE \quad \quad \quad \quad $P_{G(i)}^{t}=P_{G(i)}^{t-1}+UR_{G(i)}$, $\Delta U_{i} = \Delta P_{G(i)}^t-UR_{G(i)}$. 
\STATE \quad \quad \quad \quad Dispatch $\Delta U_{i}$ in the order of $PL$
\STATE \quad \quad \quad \quad subject to ramp and generation constraints.
\STATE \quad \quad \quad \textbf{End if} 
\STATE \quad \quad \textbf{End for}
\STATE \quad \textbf{End for}
\STATE \textbf{End} 
\end{algorithmic}
\end{algorithm}

\subsubsection{Flow charts with the proposed algorithm}
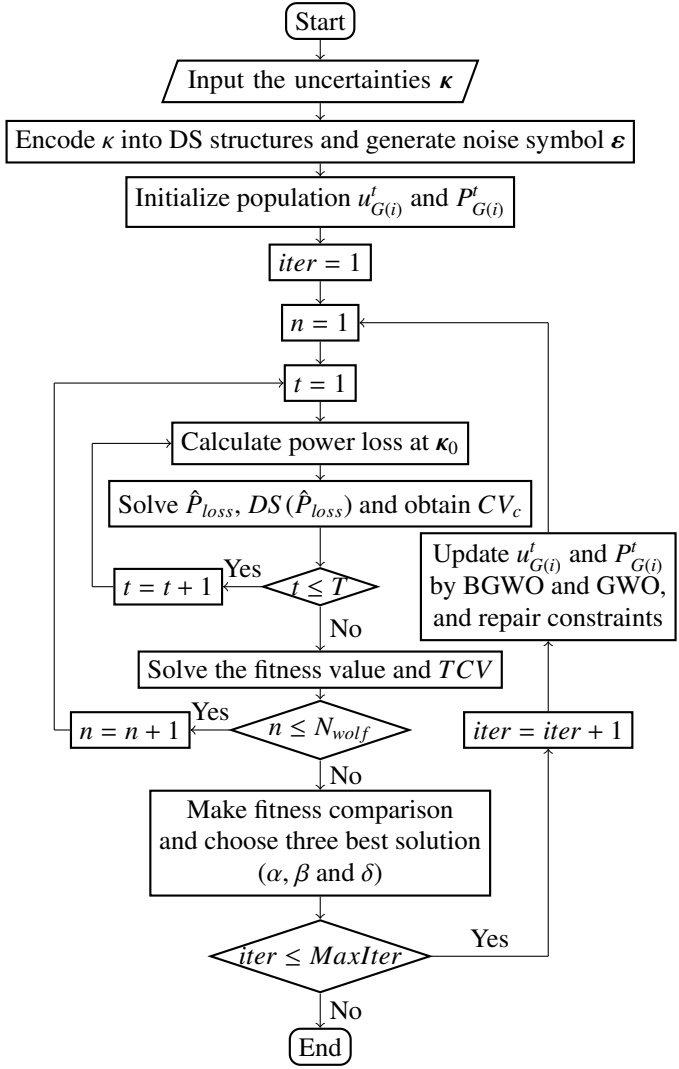
\begin{figure}[!ht] 
\centering
\begin{adjustbox}{max width=0.8\textwidth}   
\tikzset{
decision/.style={diamond, draw, thick, aspect=3, text badly centered, inner sep=0pt},
block/.style={rectangle, draw, thick, text centered, rounded corners},
process/.style={rectangle, draw, thick, text centered},
input/.style={trapezium, trapezium left angle=70, trapezium right angle=-70, draw, thick, text width=10em, text centered},
point/.style={coordinate,on grid}  
}
\begin{tikzpicture}[node distance=8mm]
\node[block] (start){Start};
\node[input,below of=start,node distance=8mm] (input){Input the uncertainties $\boldsymbol{\kappa}$};
\node[process,below of=input,node distance=8mm] (pro1){Encode $\kappa$ into DS structures and generate noise symbol $\boldsymbol{\varepsilon}$};
\node[process,below of=pro1,node distance=8mm] (pro2){Initialize population $u_{G(i)}^t$ and $P_{G(i)}^t$};
\node[process,below of=pro2,node distance=8mm] (pro3){$iter=1$};
\node[process,below of=pro3,node distance=8mm] (pro4){$n=1$};
\node[point, right of=pro4, node distance=30mm](pointiter2){};
\node[process,below of=pro4] (pro5){$t=1$};
\node[point, left of=pro5, node distance=35mm](pointwolf2){};

\node[process,below of=pro5] (pro6){Calculate power loss at $\boldsymbol{\kappa}_{0}$};
\node[point, left of=pro6, node distance=30mm](pointtime2){};
\node[process,below of=pro6] (pro7){Solve $\hat{P}_{loss}$, $DS(\hat{P}_{loss})$ and obtain $CV_c$};
\node[decision,below of=pro7,node distance=11mm] (decision1){$t\le T$};
\node[process,right of=decision1, node distance=30mm, align=center] (update){Update $u_{G(i)}^t$ and $P_{G(i)}^t$ \\by BGWO and GWO,\\and repair constraints};
\node[process, left of=decision1, node distance=20mm](judgetime){$t=t+1$};
\node[point, left of=judgetime, node distance=10mm](pointtime1){};
\node[process,below of=decision1,node distance=11mm] (pro8){Solve the fitness value and $TCV$};
\node[decision,below of=pro8] (decision2){$n\le N_{wolf}$};
\node[process, left of=decision2, node distance=25mm](judgewolf){$n=n+1$};
\node[point, left of=judgewolf, node distance=10mm](pointwolf1){};
\node[process, right of=decision2, node distance=30mm](judgeiter){$iter=iter+1$};
\node[process,below of=decision2,align=center,node distance=15mm] (pro9){Make fitness comparison \\and choose three best solution \\ ($\alpha$, $\beta$ and $\delta$)};
\node[decision,below of=pro9,node distance=15mm] (decision3){$iter\le MaxIter$};
\node[point, right of=decision3, node distance=30mm](pointiter1){};
\node[block,below of=decision3,node distance=12mm] (end){End};
\draw[->] (start)--(input);
\draw[->] (input)--(pro1);
\draw[->] (pro1)--(pro2);
\draw[->] (pro2)--(pro3);
\draw[->] (pro3)--(pro4);
\draw[->] (pro4)--(pro5);
\draw[->] (pro5)--(pro6);
\draw[->] (pro6)--(pro7);
\draw[->] (pro7)--(decision1);
\draw[->] (decision1)--node[right]{No}(pro8);
\draw[->] (pro8)--(decision2);
\draw[->] (decision2)--node[right]{No}(pro9);
\draw[->] (pro9)--(decision3);
\draw[->] (decision3)--node[right]{No}(end);

\draw[->] (decision1)--node[above]{Yes}(judgetime);
\draw[-]  (judgetime)--(pointtime1);
\draw[-]  (pointtime1)--(pointtime2);
\draw[->] (pointtime2)--(pro6.west);

\draw[->] (decision2)--node[above]{Yes}(judgewolf);
\draw[-]  (judgewolf)--(pointwolf1);
\draw[-]  (pointwolf1)--(pointwolf2);
\draw[->] (pointwolf2)--(pro5.west);

\draw[-] (decision3)--node[above]{Yes}(pointiter1);
\draw[->]  (pointiter1)--(judgeiter);
\draw[->]  (judgeiter)--(update);
\draw[-]  (update)--(pointiter2);
\draw[->] (pointiter2)--(pro4.east);
\end{tikzpicture}
\end{adjustbox}
\caption{Flow chart of UC solution by using the enhanced GWO algorithm.}
\label{fig-algorithm}
\end{figure}

The flow chart of the UC solution by using the enhanced GWO algorithm is given in~\autoref{fig-algorithm}.


\section{Experiment analysis} 
\label{sec:experiment_analysis}
In this section, the standard IEEE 30-bus system and a real-sized 183-bus China power system are employed to test the proposed model and algorithm. The probability distribution of wind speed is assumed to be Weibull distribution, and the corresponding shape and scale parameters $k$ and $\lambda$ are set to 2.49 and 6.85m/s, respectively. The CDF of wind farm without considering wake effects is represented as follows:
\begin{equation}
F(nP_{wt})=\\
\begin{cases}
0, \quad \quad \quad \quad \quad \quad \quad \quad nP_{wt} < 0\\
-exp\left\{-\dfrac{1}{(\lambda)^k}[\dfrac{nP_{wt}}{nP_{wtr}}(v_r^3-v_{ci}^3)+v_{ci}^3]^{k/3}\right\}\\ 
\quad +exp\left[-(\dfrac{v_{co}}{\lambda})^k\right]+1, 0 \le nP_{wt} < nP_{wtr}\\
1, \quad \quad \quad \quad \quad \quad \quad \quad nP_{wt} \ge nP_{wtr}
\end{cases}
\label{equ-cdf-wind}
\end{equation}
where $n$ is the number of wind turbines, $v_r$, $v_{ci}$ and $v_{co}$ mean the rated wind speed, cut-in wind speed and cut-out wind speed, and $P_{wt}$ and $P_{wtr}$ are respectively represented as the output power and the rated power of wind turbine.

Besides, we consider that all uncertain factors are independent, and the central value of load in each bus is directly proportional to the total load central value. The DS structure in our experiments is composed of 100 equiprobable focal elements. The pre-set thresholds $\sigma_c$ given in inequality constraints with DS structure for power balance, spinning reserve, and network security are set to 0.9, 1, and 1, respectively, and the whole scheduling period is 24 hours. All simulations are implemented under the MATLAB$^{TM}$ environment on an Intel Core i5-4460 CPU and 8 GB RAM personal computer.


\subsection{Case 1: IEEE 30-bus system} 
In the IEEE 30-bus system, a wind farm is connected to bus 6, and the sum of load demand is listed in \autoref{load-demand-ieee30} during the whole scheduling period. The types of uncertain inputs are shown in \autoref{uncertain-description-ieee30}, consisting of probability distributions, intervals, and fuzzy numbers, and the numerical values are given in \autoref{uncertain-value-ieee30}. In our work, we assume that the load at each bus at hour $t$ and the uncertain load inputs are proportional to the sum of load demand.

The penalty cost $\xi_{p}$ in the objective function and $\Delta$ in power balance constraints are preset to 0.1 and 0.1MW, respectively. The population size of grey wolves is set to 100, and the maximum number of iteration is 500.
\begin{table}[!ht]
\centering\small
\setlength{\abovecaptionskip}{2pt}
\caption{Load demand in the IEEE 30-bus system}
\begin{tabular*}{0.5\textwidth}{cccccc}
\toprule
t & $\sum_{b\in NB}P_{d,b}^{t}$ & t& $\sum_{b\in NB}P_{d,b}^{t}$ & t& $\sum_{b\in NB}P_{d,b}^{t}$\\
(h) & (MW) & (h) & (MW) & (h) & (MW)\\
\midrule
1 & 190.54 & 9  & 246.49 & 17 & 226.01  \\
2 & 204.58 & 10 & 244.44  &18 & 214.49 \\
3 & 203.99 & 11 & 285.65  &  19 & 222.51 \\
4 & 205.78 & 12 & 285.21 &20 & 206.32 \\
5 & 207.81 & 13 & 283.54 & 21 & 211.50 \\
6 & 212.52 & 14 & 281.30 &  22 & 246.13  \\
7 & 237.65 & 15 & 284.13 & 23 & 246.05   \\
8 & 252.44 & 16 & 250.17 & 24 & 216.23 \\
\bottomrule
\end{tabular*}
\label{load-demand-ieee30}
\end{table}
\begin{table}[!ht]
\centering\small
\setlength{\abovecaptionskip}{2pt}
\caption{Uncertain inputs in the IEEE 30-bus system}
\begin{tabular*}{0.5\textwidth}{cccc}
\toprule
Input & Bus & Mathematical model & Detail\\
\midrule
$P_{W(1)}$		& 6  & Probability distribution & Weibull distribution\\
$P_{d,12}$		& 12 & Interval & / \\ 
$P_{d,21}$		& 21 & Possible distribution & Triangular fuzzy number \\
\bottomrule
\end{tabular*}
\label{uncertain-description-ieee30}
\end{table}
\begin{table}[!ht]
\centering\small
\setlength{\abovecaptionskip}{2pt}
\begin{threeparttable}[b]
\caption{Wind farms and loads at different levels in the IEEE 30-bus system}
\begin{tabular*}{0.5\textwidth}{ccccc}
\toprule
Input & Bus & Wind & Load & Power \\
	  & 	& penetration &  deviation  & (Unit: MW) \\
\midrule
\multirow{3}{*}{$P_{W(1)}$}		& \multirow{3}{*}{6} & 10\% & / & 28\tnote{1}\\
 & & 20\% & / & 56\tnote{1}\\
& & 30\% & / & 84\tnote{1}\\
\midrule
\multirow{3}{*}{$P_{d,12}$}		& \multirow{3}{*}{12} & / & 10\% & [10.08,12.32]\\
 & & / & 15\% & [9.52,12.88] \\
& & / & 20\% & [8.96,13.44]\\
\midrule
\multirow{3}{*}{$P_{d,21}$}		& \multirow{3}{*}{21} & / & 10\% & (15.75,17.50,19.25)\\
 & & / & 15\% & (14.87,17.50,20.12)\\
& & / & 20\% & (14.00,17.50,21.00)\\
\bottomrule
\end{tabular*}
\begin{tablenotes}
     \item[1] The power value here means the rated power of wind farm.
\end{tablenotes}
\label{uncertain-value-ieee30}
\end{threeparttable}
\end{table}
 
Aiming to conduct the algorithm comparison, three scenarios are considered for the simulations: 1) traditional GWO; 2) traditional GWO with initial population optimization; 3) the enhanced GWO algorithm. The levels of load deviation are all set to 15\%, and the corresponding results pertinent to three scenarios (the values of $TCV$) are shown in \autoref{result-algorithm}. Apparently, the values of $TCV$ based on the enhanced GWO algorithm in three different penetration levels are lower than the values in the other two scenarios, demonstrating a better performance of the proposed algorithm.  
\begin{table}[!ht]
\small
\setlength{\abovecaptionskip}{2pt}
\caption{The values of $TCV$ in three scenarios}
\begin{tabular*}{0.5\textwidth}{p{2.5cm}p{1.5cm}<{\centering}p{1.5cm}<{\centering}p{1.5cm}<{\centering}}
\toprule
Wind penetration & 10\%  & 20\%  & 30\% \\
\midrule
Scenario 1	& 19.95 & 20.93 & 20.95 \\
Scenario 2	& 19.70 & 19.93 & 19.78 \\
Scenario 3	& 19.45 & 19.58 & 19.76 \\
\bottomrule
\end{tabular*}
\label{result-algorithm}
\end{table}

\begin{figure}[!ht]
\centering
\includegraphics[width=0.5\textwidth]{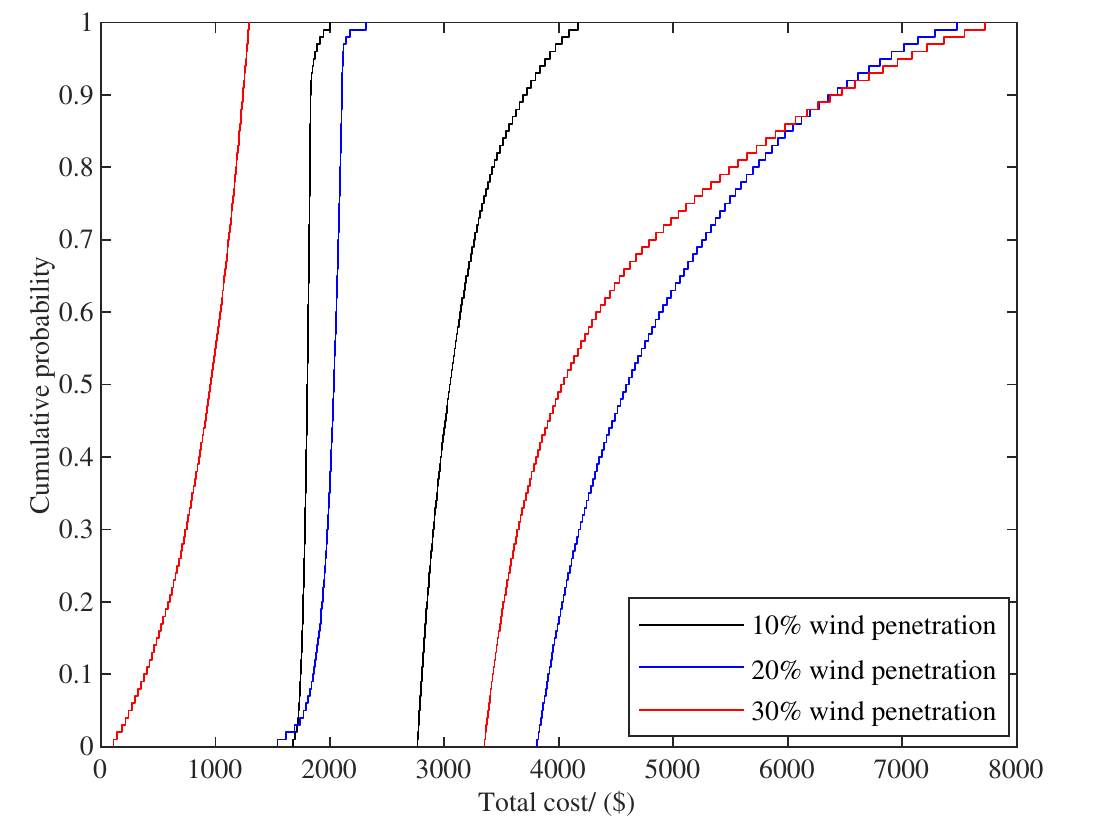}
\caption{P-boxes of total cost in IEEE 30-bus system at different levels of wind penetration.}
\label{fig-plot_ieee30_pen}
\end{figure}

\begin{figure}[!ht]
\centering
\includegraphics[width=0.5\textwidth]{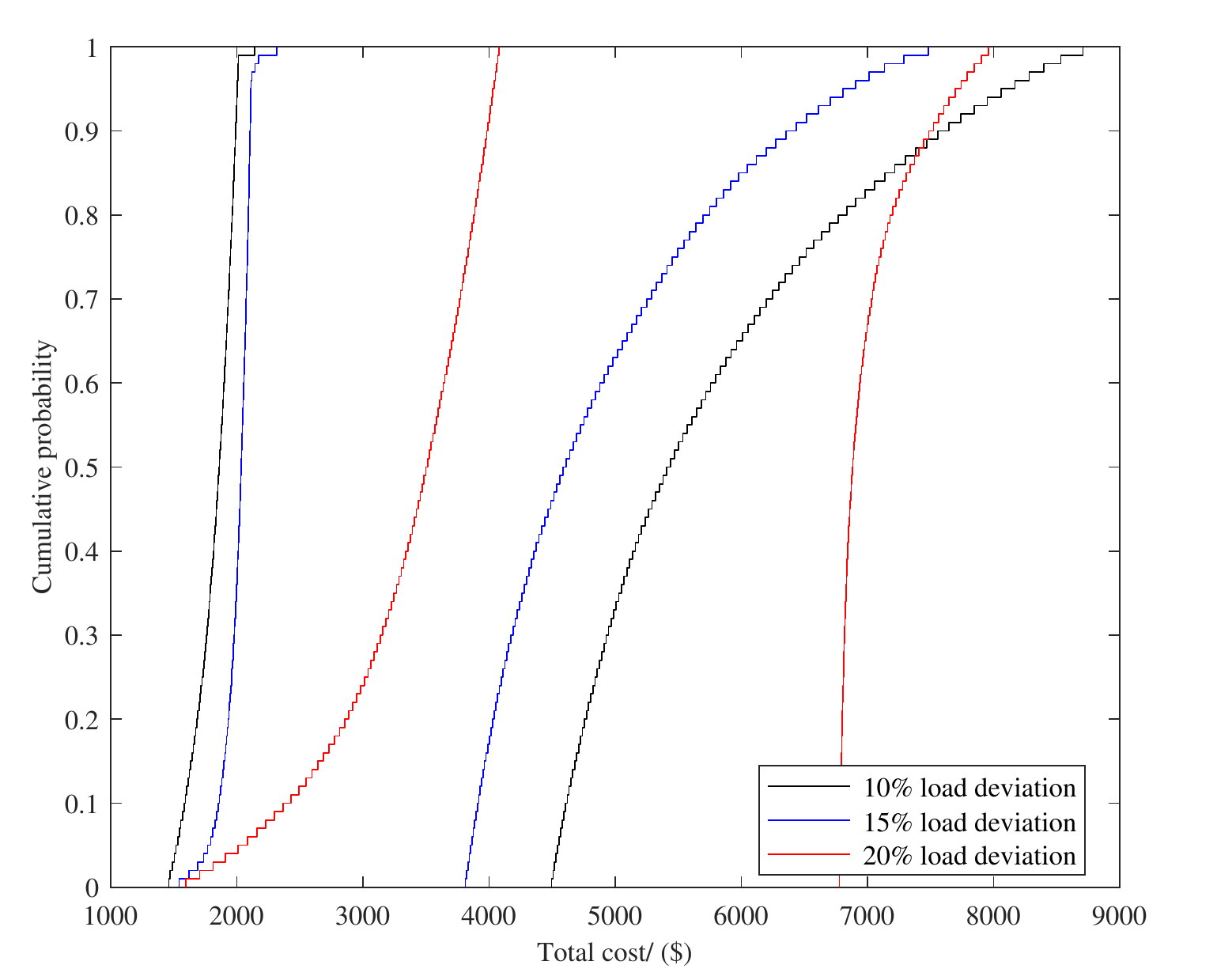}
\caption{P-boxes of total cost in IEEE 30-bus system at different levels of load deviation.}
\label{fig-plot_ieee30_load}
\end{figure}

The total costs at different wind power penetration levels, namely 10\%, 20\%, and 30\% are shown in \autoref{fig-plot_ieee30_pen} under 15\% load deviation level. From \autoref{fig-plot_ieee30_pen}, it can be seen that the range of P-box is expanding as the level of wind power penetration increases. What's more, the CDFs of right bounds are rapidly moving to the right side compared to the left bounds under high-level wind power penetration, reflecting that the worst-case in wind generation does greatly increase the scheduling cost. In other words, if uncertain power generation still remains a high penetration level, the potential cost increase in the worst-case is much greater than the cost reduction from consuming wind power. By comparison, \autoref{fig-plot_ieee30_load} shows that the CDFs of left and right bounds are both going to the right side with the different levels of load deviation under 20\% wind power penetration level, however, there are no significant changes in the range of P-box. The possible reason is that we apply the fuzzy numbers and intervals to express the uncertainty of load, which brings epistemic uncertainty during optimization. It can be also obtained from the simulation results that the increase in scheduling costs in \autoref{fig-plot_ieee30_pen} and \autoref{fig-plot_ieee30_load} also reminds us of the significance of reducing load uncertainty and improving the accuracy of the load forecast.

\begin{figure}[!ht]
\centering
\includegraphics[width=0.5\textwidth]{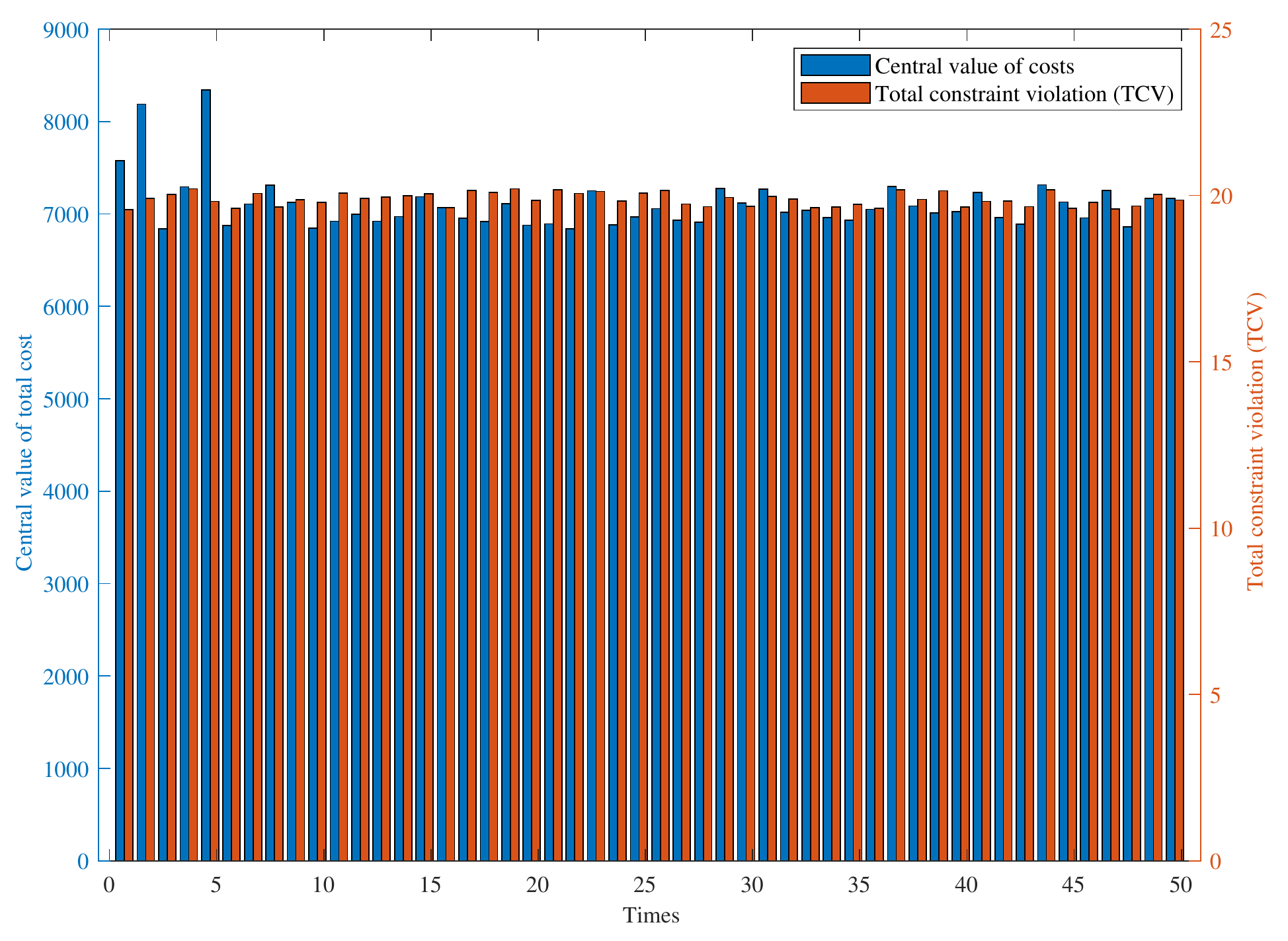}
\caption{Results of the central value of total costs and total violations by executing 50 trials.}
\label{fig-plot_stability}
\end{figure}

The stability of the proposed algorithm is also studied in Case 1 by executing 50 trials with 20\% wind power penetration level and 15\% load deviation. \autoref{fig-plot_stability} provides the optimization results of the central value of total costs and total violations. The standard deviations of the central value of total costs and total violations are 312.12 and 0.198, respectively. It can be stated that the enhanced GWO algorithm has good convergence and stability.

\label{sub:case_1_modified_ieee_30_bus_system}

\subsection{Case 2: real-sized 183-bus China power system} 
\label{sub:case_2_real_large_scale_183_bus_system}
\begin{figure}[!ht]
\centering
\includegraphics[width=0.5\textwidth]{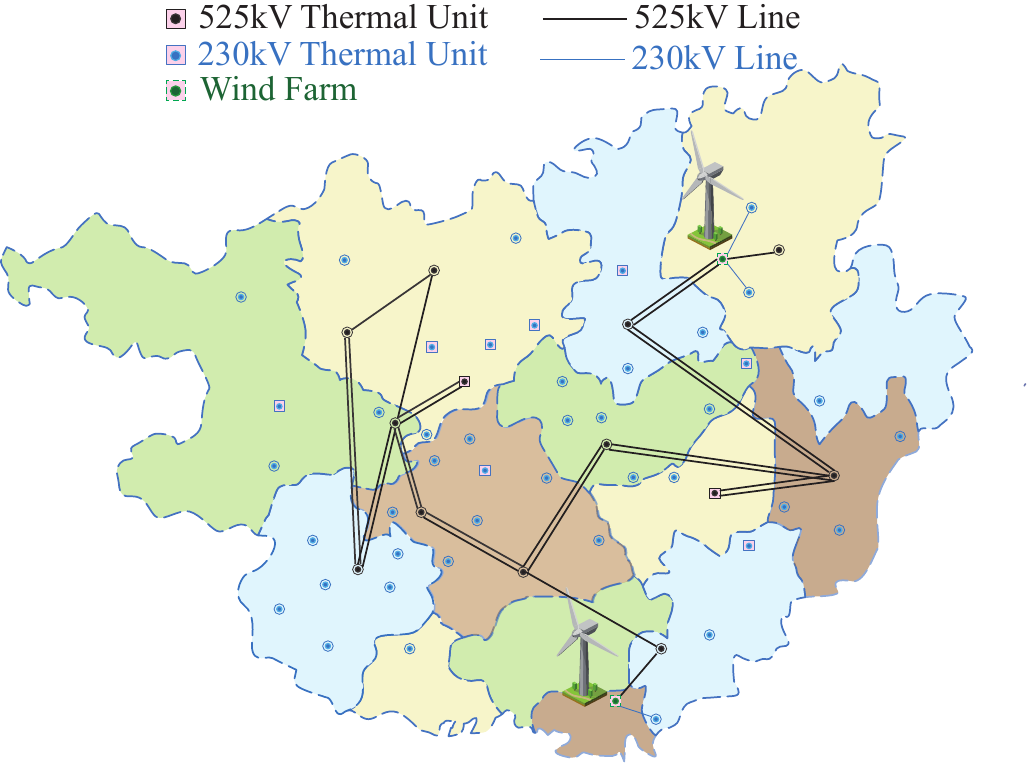}
\caption{Geographic wiring diagram of real-sized 183-bus China power system.}
\label{fig-plot_jx183_geography}
\end{figure}

We also test the proposed model and algorithm in a real-sized 183-bus China power system, which is composed of 183 buses, 308 branches, and 30 traditional generation units~\cite{luo2018solution}. The geographic wiring diagram is shown in \autoref{fig-plot_jx183_geography}. Two wind farms are located at bus 69 and bus 156. The types of uncertain inputs and the details of uncertain factors are listed in \autoref{uncertain-description-jx183} and \autoref{uncertain-value-jx183}, respectively. The sum of peak load demands is set to 8626.81MW, and $\Delta$ is preset to 1 MW.
\begin{table}[!ht]
\centering\small
\setlength{\abovecaptionskip}{2pt}
\caption{Uncertain inputs in a real-sized 183-bus China power system}
\begin{tabular*}{0.5\textwidth}{cccc}
\toprule
Input & Bus & Mathematical model & Detail\\
\midrule
$P_{W(1)}$		& 69  & Probability distribution & Weibull distribution\\
$P_{W(2)}$		& 156  & Probability distribution & Weibull distribution\\
$P_{d,108}$		& 108 & Interval & / \\ 
$P_{d,72}$		& 72 & Possible distribution & Triangular fuzzy number \\
\bottomrule
\end{tabular*}
\label{uncertain-description-jx183}
\end{table}

\begin{table}[!ht]
\centering\small
\setlength{\abovecaptionskip}{2pt}
\begin{threeparttable}[b]
\caption{Wind farms and loads at different levels in real-sized 183-bus China power system}
\begin{tabular*}{0.5\textwidth}{ccccc}
\toprule
Input & Bus & Wind & Load & Power \\
	  & 	& penetration &  deviation  & (Unit: MW) \\
\midrule
\multirow{3}{*}{$P_{W(1)}$}		& \multirow{3}{*}{69} & 10\% & / & 430\tnote{1}\\
 & & 20\% & / & 862\tnote{1}\\
& & 30\% & / & 1294\tnote{1}\\
\midrule
\multirow{3}{*}{$P_{W(2)}$}		& \multirow{3}{*}{156} & 10\% & / & 432\tnote{1}\\
 & & 20\% & / & 864\tnote{1}\\
& & 30\% & / & 1294\tnote{1}\\
\midrule
\multirow{3}{*}{$P_{d,108}^{11}$}		& \multirow{3}{*}{108} & / & 10\% & [96.07,117.41]\\
 & & / & 15\% & [90.73,122.75] \\
& & / & 20\% & [85.39,128.09]\\
\midrule
\multirow{3}{*}{$P_{d,72}^{11}$}		& \multirow{3}{*}{72} & / & 10\% & (206.66,229.62,252.58)\\
 & & / & 15\% & (195.18,229.62,264.06) \\
& & / & 20\% & (195.18,229.62,264.06)\\
\bottomrule
\end{tabular*}
\begin{tablenotes}
     \item[1] The power value here means the rated power of wind farm.
   \end{tablenotes}
\label{uncertain-value-jx183}
\end{threeparttable}
\end{table}

\begin{figure}[!ht]
\centering
\includegraphics[width=0.5\textwidth]{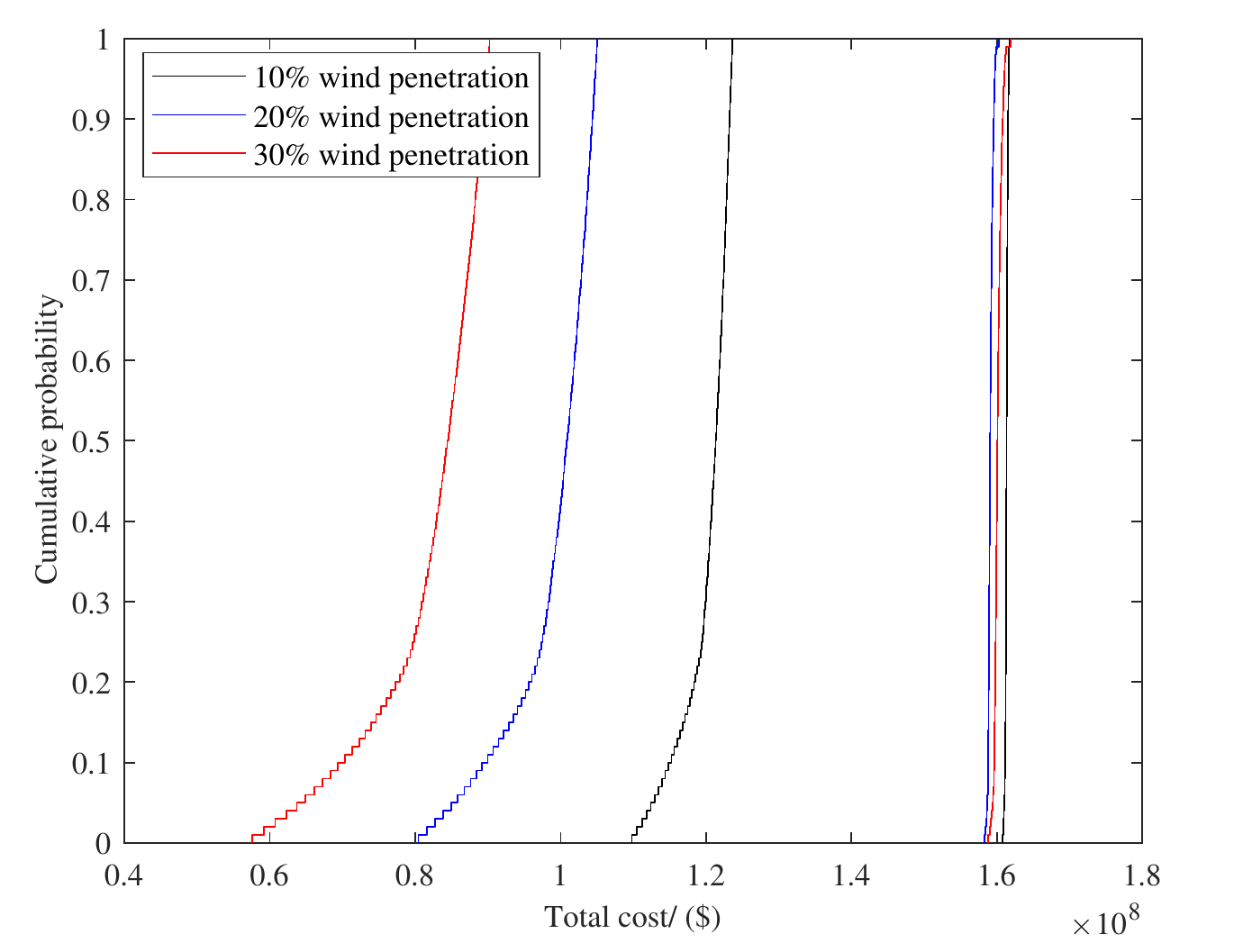}
\caption{P-boxes of total cost in real-sized 183-bus China power system at different levels of wind penetration.}
\label{fig-plot_jx183_pen}
\end{figure}

\begin{figure}[!ht]
\centering
\includegraphics[width=0.5\textwidth]{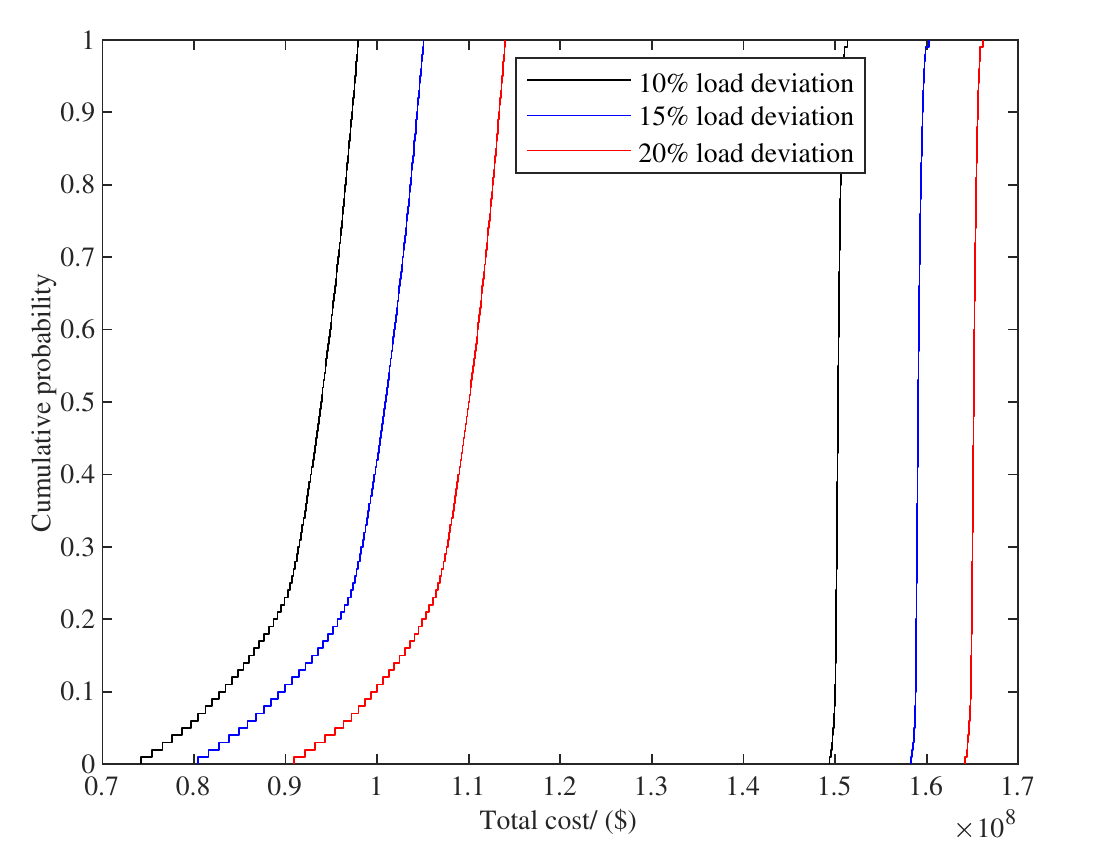}
\caption{P-boxes of total cost in real-sized 183-bus China power system at different levels of load deviation.}
\label{fig-plot_jx183_load}
\end{figure}

Similar to Case 1,~\autoref{fig-plot_jx183_pen} depicts the total cost at different levels of wind power penetration under 15\% load deviation level, and \autoref{fig-plot_jx183_load} shows the total cost at different levels of load deviation with 20\% wind power penetration level. On one hand, along with the increasing level of wind power penetration, the CDF of left bound moves to the left side, but the right bound basically remains unchanged, showing that the costs in the worst-case wind generation are not greatly influenced by the level of wind power penetration in this real-sized 183-bus China power system. On the other hand, the shape and the moving trend of costs under different load deviation levels match the results from the previous case.

\section{Conclusions and Future Discussion} 
\label{sec:conclusions_and_future_discussion}
A novel approach for the solution of UC problem incorporating manifold uncertainties is proposed in this paper. By utilizing the hybrid ET and EAA approach, the probability, possibility, and interval measures, described as wind generation and load deviation, can be fully considered in UC problem to demonstrate the aleatory and epistemic uncertainties of scheduling costs in the form of P-boxes. The uncertainty of power loss is also introduced to obtain the more practical UC optimization results, and an enhanced GWO algorithm is applied for solving this problem. Besides, some key treatments for UC with manifold uncertainties are introduced, containing fitness comparison, inequality constraints with DS structure, initial population optimization and constraints repair. After applying to the IEEE 30-bus system and a real-sized 183-bus China power system, the experiment analyses demonstrate the validity and scalability of the proposed model and method. Future researches may focus on the consideration of the correlation between uncertainties with multiple wind farms in UC optimization problems.

\section*{Acknowledgements}
This work was supported in part by the National Natural Science Foundation of China (51777103).

\bibliography{mybibfile}

\end{document}